\documentclass{elsarticle}

\usepackage{lineno,hyperref}
\usepackage{amsmath,amssymb,color,latexsym,natbib,graphicx}
\usepackage{dcolumn}
\usepackage{ulem}
\usepackage{bm}
\usepackage{yfonts}
\usepackage{calrsfs}
\DeclareMathAlphabet{\pazocal}{OMS}{zplm}{m}{n}
\def\ADD#1{{\textcolor{black}{#1}}}    
\newcommand{\p} {\partial}
 
\def\ie{{\it i.e.}\ }
\def\uu{{\bf u}}
\def\bomega{{\bf w}}

\def\d_M{{\bf d_M}}
\def\fz{{\bf f_{\bf u}}}

\def\jj{{\bf j}}
\def\bb{{\bf b}}

\def\xx{{\bf x}}

\def\bOmega{{\boldsymbol \Omega}}
\def\bomega{{\boldsymbol \omega}}
\def\be{\begin{equation}}
\def\ee{\end{equation}}
\def\ba{\begin{eqnarray}}
\def\ea{\end{eqnarray}}
\def \pmbmath{\mathpalette\pmbmathaux}
\def \pmbmathaux#1#2{
         \pmbtext{$#1#2$}}
\def \pmbtext#1{\leavevmode
     \setbox0\hbox{#1}
     \kern0,4pt \copy0 \kern-\wd0
     \kern-0,2pt \raise0,3pt \box0 }
\def\ellb{\pmbmath{\ell}}
\def\nablab{\pmbmath{\nabla}}

\modulolinenumbers[5]
\journal{Journal of \LaTeX\ Templates}
\begin{document}
\begin{frontmatter}
\title{An alternative formulation for exact scaling relations in hydrodynamic and magnetohydrodynamic turbulence}
\author[mymainaddress]{Supratik Banerjee\corref{mycorrespondingauthor}}
\cortext[mycorrespondingauthor]{Corresponding author}
\ead{supratik.banerjee@uni-koeln.edu}
\author[mysecondaryaddress,mythirdaddress]{S\'ebastien Galtier}
\address[mymainaddress]{Institut f\"ur Geophysik und Meteorologie, Universit\"at zu K\"oln, Germany}
\address[mysecondaryaddress]{Laboratoire de Physique des Plasmas, \'Ecole polytechnique, Palaiseau, France}
\address[mythirdaddress]{Universit\'e Paris-Sud, Orsay, France}

\begin{abstract}
We propose an alternative formulation for the exact relations in three-dimen\-sional homogeneous turbulence using two-point statistics. 
Our finding is illustrated with incompressible hydrodynamic, standard and Hall magnetohydrodynamic turbulence. In this formulation, 
the cascade rate of an inviscid invariant of turbulence can be expressed simply in terms of mixed second-order structure functions. 
Besides the usual variables like the velocity $\uu$, vorticity $\bomega$, magnetic field $\bb$ and the current $\jj$, the vectors 
$\uu \times \bomega$, $\uu \times \bb$ and $\jj \times \bb$ are also found to play a key role in the turbulent cascades. 
The current methodology offers a simple algebraic form which is specially interesting to study anisotropic space plasmas like the solar wind, 
\ADD{with in principle a faster statistical convergence than the classical laws written in terms of third-order correlators.}
\end{abstract}


\end{frontmatter}

\linenumbers

\paragraph*{Introduction} \label{intro}
Turbulence is said to be the last unsolved problem of classical mechanics. 
Despite being ubiquitous in nature, the highly non-linear character of turbulence renders it challenging to be studied analytically.
Most of the theoretical studies of turbulence is based on intuitive phenomenological ideas and numerical simulations. 
There exists only a few analytical relations which are derived analytically by using a two-point statistics. These relations are called 
exact relations and they are certainly among the most important elements in the statistical theory of strong turbulence. Exact relations 
express the mean transfer rate (denoted by $\varepsilon$) of an inviscid invariant of turbulence (e.g. kinetic energy in incompressible hydrodynamics, 
cross-helicity in incompressible magnetohydrodynamics (MHD))
in terms of the length scale and the statistical moments of two-point fluctuations of different variables (e.g. velocity, magnetic field etc.). 
Unlike the phenomenological predictions, these analytical relations give an accurate measure of $\varepsilon$ which characterises the universality 
of a turbulent system in physical space.

The simplest result in this domain is certainly the pioneering four-fifths law of Kolmogorov 
\cite{Kolmogorov41,Frisch95} (see also \cite{Antonia1997} for the four-thirds law) for incompressible hydrodynamics which reads
\begin{equation}
\left\langle \delta u_\ell^3 \right\rangle = - \frac{4}{5} \varepsilon \ell \, ,
\label{firste}
\end{equation}  
where $\delta X$ denotes the difference of a physical quantity $X$ (here the fluid velocity $\uu$) between two arbitrary points $\xx$ and 
$\xx' \equiv \ \xx + \ellb$, $\ell = \vert \ellb \vert$, $\delta u_\ell = \delta \uu \cdot {\bm \ell/ \ell}$, $\left\langle \cdot \right\rangle$ stands for 
an ensemble average and $\varepsilon$ 
is the mean cascade rate of energy (which is also equal to the mean energy dissipation rate (per unit mass) for stationary turbulence). 
This law describes a universal property of a three-dimensional, homogeneous, 
isotropic turbulence in the physical space and in the limit of vanishing kinematic viscosity.
Later, Monin and Yaglom \cite{Monin75} proposed a differential form of the above relation without using the isotropy assumption, \ie
\begin{equation}
\nablab_{\ellb} \cdot \left\langle \delta u^2 \delta \uu \right\rangle = - 4 \varepsilon \, . \label{vec1}
\end{equation}
This type of statistical law has also been derived for passive scalars \cite{Yaglom49}, or conducting MHD fluids 
\cite{Politano98,Politanob98} using both tensorial and vectorial formalisms corresponding to the total energy and cross-helicity conservation. 
In recent years, a number of theoretical efforts have been dedicated to the derivation of exact statistical relations for astrophysical or 
space plasmas starting from incompressible Hall MHD, electron MHD, to compressible (supersonic) hydrodynamics and MHD 
\cite{Galtier08H,Galtier08,Meyrand10,Galtier11,Banerjee13}. 
Following the tensorial formalism of von K\'arm\'an \& Howarth, isotropic exact laws have also 
been obtained for helical turbulence both in hydrodynamics and MHD \cite{Chkhetiani96,Gomez00,Politano03}, 
but the final relations are not expressible exclusively in terms of two-point fluctuations 
(\ie increments) and therefore cannot readily be amenable to spectral prediction or phenomenological interpretation. 
However, using vector algebra, we have recently derived helical exact relations for Hall MHD \cite{Banerjee16} which can be 
expressed purely in terms of two-point fluctuations.

In the present paper, we follow the same methodology as in \cite{Banerjee16} to derive various exact relations for some inviscid 
invariants in three-dimensional homogeneous incompressible hydrodynamic, standard and Hall MHD turbulence \ADD{in the limit of infinitely large 
kinetic and magnetic Reynolds numbers ${\textgoth R_e}$ and ${\textgoth R_m}$ respectively.} 
We also discuss the relevance of these new expressions for space plasmas where the assumption of statistical isotropy does not hold.

\section{Incompressible hydrodynamic turbulence}
\subsection{Conservation of energy}
We start with an alternative form \cite{Lamb1878} of Navier-Stokes equations
\begin{align}
\p_t{\uu} &= -\nabla P_T  + \uu \times {\boldsymbol \omega} + \nu \nabla^2 \uu  + \fz \, , \label{lamb} \\
\nabla \cdot \uu &= 0 \, , 
\end{align}
where $P_T = P + {u^2}/{2}$, P being the fluid pressure, $\bomega = \nabla \times \uu$ the vorticity vector, $\nu$ the coefficient of kinematic viscosity and 
$\fz$ represents a large-scale stationary forcing.
In this section, we shall derive an exact relation corresponding to the conservation of the kinetic energy $E_K= \int {u^2}/2 \ d \tau$, $\tau$ being the volume of integration. 
As usual \cite{Batchelor}, we will implicitly assume that $\uu$ is a fluctuating velocity field (\ie $\langle \uu \rangle = 0$).
The two-point (symmetric) correlators for the kinetic energy  are defined as
\begin{equation}
R_{E} = R'_{E} = \left\langle \frac{\uu \cdot \uu'}{2} \right\rangle \, , 
\label{corre1}
\end{equation}
where the prime denotes variables at point $\xx'$ (see the comment after Eq. (\ref{firste})). 
The evolution equation of the correlators is then given by
\begin{align}
\p_t  &\left( R_{E} + R'_{E} \right) = \left\langle \uu' \cdot \p_t \uu + \uu \cdot \p_t \uu'   \right\rangle \nonumber \\
&= \left\langle \uu' \cdot \left[ - \nabla P_T + \uu \times{\boldsymbol  \omega} + \nu \Delta \uu + \fz \right]\right\rangle 
+ \left\langle \uu \cdot \left[ - \nabla' P'_T + \uu' \times{\boldsymbol  \omega'} + \nu \Delta' \uu' + \fz' \right]\right\rangle \nonumber \\
&= \left\langle \uu' \cdot \left( \uu \times{\boldsymbol  \omega} \right) + \uu \cdot \left( \uu' \times{\boldsymbol  \omega'} \right) \right\rangle 
+ {\cal D} + {\cal F} \, , 
\label{ns1}
\end{align}
where ${\cal D} = \nu \left\langle \uu \cdot \Delta' \uu' + \uu' \cdot \Delta \uu \right\rangle$ and ${\cal F} = \left\langle \uu \cdot \fz' + \uu' \cdot \fz \right\rangle$. By incompressibility and statistical homogeneity, we get
\begin{align}
\left\langle \uu \cdot \nabla' P_T' \right\rangle = \nabla_{\ellb} \cdot \left\langle P_T' \uu \right\rangle 
= - \left\langle P_T' \left( \nabla \cdot \uu \right) \right\rangle = 0
\end{align}
and similarly $\left\langle \uu' \cdot \nabla P_T \right\rangle = 0$, hence the form (\ref{ns1}). 
Now, we consider a statistical stationary state for which
$$
\p_t   \left( R_{E} + R'_{E} \right) = 0 \, . 
$$
For length-scales well inside the inertial range, we can neglect the dissipative term and obtain 
\begin{align}
\left\langle \left( \uu \times {\boldsymbol  \omega} \right) \cdot \uu' + \left( \uu' \times {\boldsymbol  \omega}' \right) \cdot \uu \right\rangle 
= - 2 \varepsilon \, ,
\end{align}
where ${\cal F} \simeq 2 \varepsilon$ (large-scale forcing). After simple manipulations, we obtain the final form of the exact relation 
\begin{equation}
\left\langle  \delta (\uu \times {\boldsymbol \omega}) \cdot \delta  \uu  \right\rangle =  2 \varepsilon \, , \label{first}
\end{equation}
where we have used the relations $\left( \uu \times {\boldsymbol \omega} \right) \cdot \uu = 0$ and 
$\left( \uu' \times {\boldsymbol \omega}' \right) \cdot \uu' = 0$. 

Eq. (\ref{first}) is the first result of our paper. 
This gives a divergence free exact relation for homogeneous incompressible turbulence without any prior assumption of statistical isotropy. 
Unlike the four-fifths law (\ref{firste}) or its differential form (\ref{vec1}), the new expression does not involve a third-order structure function but a 
mixed second-order structure function. 
\ADD{Hence, the evaluation of $\varepsilon$ becomes easier since the statistical convergence is expected to be faster for lower order moment \cite{Podesta09}.}
In this case an estimation of $\varepsilon$ can be obtained directly from the measurement of the scalar product of the fluctuations of the 
Lamb vector, $- (\uu \times  \bomega)$, with the velocity. 
From this new divergence free form, \ADD{we see} that the mean turbulent energy flux vanishes when the system satisfies the Beltrami conditions
(\ie $ \uu \parallel \bomega$). 
\ADD{But this is an extreme situation which demands to satisfy the alignment condition at every points of the turbulent fluid which is unlikely to happen
naturally. Note that for a state of Beltrami flow, the kinetic energy is minimum (and the kinetic helicity maximum) and so this indicates a state of stability 
thereby corresponding to zero turbulent flux.} 
We can also check the compatibility with Eq. (\ref{vec1}). Using the fact 
\begin{equation}
\left\langle \nabla' \cdot \left( \frac{u'^2}{2} \uu \right)\right\rangle  = \left\langle \nabla \cdot \left( \frac{u^2}{2} \uu' \right)\right\rangle  = 0 \, , 
\end{equation}
we find
\begin{align*}
&  \left\langle  \delta (\uu \times  \bomega) \cdot \delta  \uu  \right\rangle = - \left\langle \left( \uu \times \bomega \right) \cdot \uu' + \left( \uu' \times \bomega' \right) \cdot \uu \right\rangle \\
&=  \left\langle  \left[ \nabla' \left( \frac{u'^2}{2} \right) - \left( \uu' \times \bomega' \right)\right]  \cdot \uu + \left[  \nabla \left( \frac{u^2}{2} \right) - \left( \uu \times \bomega \right)\right]  \cdot \uu' \right\rangle \\
&=  \left\langle \left( \uu' \cdot \nabla' \right)\left( \uu' \cdot \uu \right) + \left( \uu \cdot \nabla \right)\left( \uu \cdot \uu' \right)\right\rangle =  \left\langle \nabla' \cdot \left[ \left( \uu \cdot \uu' \right) \uu' \right] + \nabla \cdot \left[ \left( \uu' \cdot \uu \right) \uu \right] \right\rangle \\
&= - \frac{1}{2} \nabla_{\ellb} \cdot \left\langle \left( \uu' \cdot \uu' - \uu \cdot \uu' - \uu' \cdot \uu + \uu \cdot \uu \right) \left( \uu' - \uu \right) \right\rangle \, .
\end{align*}
After some re-arrangements, one can recover the primitive form of the Kolmogorov's law (\ref{vec1}) for incompressible hydrodynamic 
turbulence. 
\ADD{The physical reason behind this expression is inherent to the Lamb formulation of Navier-Stokes Eq. (\ref{lamb}). 
Indeed, the average kinetic energy satisfies the following equation
\begin{equation}
\p_t \langle \frac{u^2}{2} \rangle = - \langle \nabla \cdot ( P_T \uu ) + ( \bomega \times \uu ) \cdot \uu \rangle , 
\end{equation}
which clearly shows that there is no contribution from the Lamb vector (and also from the pressure term -- after Gauss' divergence theorem). 
For the energy correlators, the situation is different because we have
\begin{equation}
\p_t \langle \frac{\uu \cdot \uu'}{2} \rangle = - \langle \nabla \cdot ( P_T \uu' ) + ( \bomega \times \uu ) \cdot \uu' \rangle .
\end{equation}
The term under the divergence still vanishes but the contribution from the Lamb vector is non-zero which shows that the turbulent energy flux rate is 
fundamentally governed by the so-called vortex force.}

An important comment has to be made about the assumption $\langle \uu \rangle = 0$ used through our derivation. Indeed, if we introduce a 
uniform velocity field ${\bf U_0}$ into expression (\ref{first}), the law is modified by the presence of an additional term, namely 
$\left\langle  \delta ({\bf U_0} \times {\boldsymbol \omega}) \cdot \delta  \uu  \right\rangle$. In fact, this new term can be understood as a contribution 
of the time derivative of the modified expression of the correlator (\ref{corre1}), which involves a scalar product between ${\bf U_0}$ and the 
velocity field. Definitively, it is not a contribution that we want to include in the description of turbulence. 
From Eq. (\ref{first}), we can easily obtain a scaling law for the Lamb vector: 
using the fundamental assumption of scale invariance of $\varepsilon$ and the relation for velocity fluctuations $\delta u \sim \ell^{1/3}$, we 
immediately find that $\vert \delta  (\uu \times \bomega)\vert  \sim \ell^{-1/3}$.

\paragraph*{Turbulence under rotation}
If the entire system is rotating with the (non uniform) angular velocity $\bOmega$, the fluid will experience an additional Coriolis acceleration 
${\mathbf{a_c}} = - 2 ( {\bOmega} \times \uu ) $. It is then sufficient to modify Eq. (\ref{first}) to 
\begin{equation}
\left\langle  \delta (\uu \times {\boldsymbol \Omega}_T) \cdot \delta  \uu  \right\rangle =  2 \varepsilon \, ,  \label{rotation1}
\end{equation}
where $\bOmega_T =  \bomega + 2\bOmega $. 
In the particular case of a solid body rotation where $\Omega$ is constant, we have 
$\left\langle  \delta (\uu \times {\boldsymbol \Omega}) \cdot \delta  \uu  \right\rangle = \left\langle (\delta \uu \times {\boldsymbol \Omega}) \cdot \delta  \uu  \right\rangle = 0$
and thus Eq. (\ref{rotation1}) simply reduces to (\ref{first}). 
It is well-known that such turbulence deviates from isotropy with a gradual columnar structuring along the rotation axis 
\cite{Cambon97,galtier09r,lamriben}. 
An estimation of $\varepsilon$ can thus be obtained directly from the measurement of the scalar product of the fluctuations of the 
Lamb vector with the velocity. This relation is a clear improvement to the model proposed by Galtier \cite{galtier09r} 
where an assumption of critical balance was introduced to integrate Eq. (\ref{vec1}) over a given manifold.

\subsection{Conservation of kinetic helicity}
In incompressible hydrodynamic turbulence, besides the kinetic energy, there is another inviscid invariant called kinetic helicity whose density is given by 
$H_K =  \uu \cdot \bomega $. The governing inviscid equations that we need to describe the evolution of this quantity are 
\begin{align}
\p_t \uu &= - \nabla P_T  + \uu \times {\boldsymbol \omega} + \nu \nabla^2 \uu  + \fz \, , \\
\p_t \bomega &= \nabla \times \left( \uu \times {\boldsymbol \omega}  \right) + \nu \nabla^2 \bomega + {\bf f}_{\bomega} \, ,
\end{align}
with ${\bf f}_{\bomega} = \nabla \times \fz$. As before, we will implicitly assume that $\uu$ is a fluctuating velocity field.
The symmetric two-point correlators for kinetic helicity are defined as
\begin{equation}
R_K = R'_K = \left\langle  \frac{\uu \cdot \bomega'  + \uu' \cdot \bomega}{2} \right\rangle \, . 
\end{equation}
Following the same line as in the previous subsection, we now calculate
\begin{align}
\p_t   &\left( R_{K} + R'_{K} \right) = \left\langle \uu' \cdot \p_t \bomega + \bomega \cdot \p_t \uu'  +  \uu \cdot \p_t \bomega' + \bomega' \cdot \p_t \uu  \right\rangle \nonumber \\
&= \left\langle \left[ \nabla \times \left( \uu \times \bomega \right) \right] \cdot \uu' + 
\left[ - \nabla' P'_T + \left( \uu' \times \bomega' \right) \right] \cdot \bomega \right\rangle \nonumber \\
&+ \left\langle \left[ \nabla' \times \left( \uu' \times \bomega' \right) \right] \cdot \uu +
\left[ - \nabla P_T + \left( \uu \times \bomega \right) \right] \cdot \bomega' \right\rangle + {\cal D}_K + {\cal F}_K \nonumber \\
&= 2 \left\langle \left( \uu \times \bomega \right) \cdot \bomega' + 
\left( \uu' \times \bomega' \right) \cdot \bomega \right\rangle + {\cal D}_K + {\cal F}_K \nonumber \\
&= - 2 \left\langle \delta \left( \uu \times \bomega \right)  \cdot \delta \bomega \right\rangle + {\cal D}_K  + {\cal F}_K \, ,
\end{align}
where the dissipation is ${\cal D}_K = \nu \left\langle \bomega \cdot \Delta' \uu' + \bomega' \cdot \Delta \uu + \uu \cdot \Delta' \bomega' 
+ \uu' \cdot \Delta \bomega \right\rangle$
and the forcing ${\cal F}_K = \left\langle \bomega \cdot \fz' + \bomega' \cdot \fz + \uu \cdot  {\bf f'}_{\bomega}     +  \uu' \cdot {\bf f}_{\bomega} \right\rangle $. In this derivation we have used the relations
\begin{equation}
\left\langle \left[  \nabla' \times \left( \uu' \times \bomega' \right)\right] \cdot \uu \right\rangle  
= \left\langle \nabla' \cdot \left[ \left( \uu' \times \bomega' \right) \times \uu \right]\right\rangle 
= \left\langle  \left( \uu' \times \bomega' \right) \cdot \bomega \right\rangle \nonumber
\end{equation}
and similarly 
$ \left\langle \left[  \nabla \times \left( \uu \times \bomega \right)\right] \cdot \uu' \right\rangle = \left\langle  \left( \uu \times \bomega \right) \cdot \bomega' \right\rangle$. 
Inside the inertial zone (in the limit of a large-scale forcing and a small viscosity), we can thus write 
\begin{equation}
\left\langle \delta \left( \uu \times \bomega \right)  \cdot \delta \bomega \right\rangle  = \varepsilon_K \, , \label{kh}
\end{equation}
where ${\cal F}_K = 2\varepsilon_K$ with $\varepsilon_K$ being the mean rate of kinetic helicity dissipation. 
Expression (\ref{kh}) is our second new exact relation. Interestingly, it gives an alternative formulation to the law found by Gomez {\it et al.} 
\cite{Gomez00} where the assumption of skew isotropy was made at the beginning of the tensorial analysis. In our case, such assumption is not made.
%

\paragraph*{Effect of background rotation}
It is straightforward to see the effect of a solid body rotation which corresponds to the substitution of $\bomega$ 
in the Navier-Stokes equations by $\bOmega_T = \bomega + 2\bOmega$, where $\bOmega$ is the rotation rate. 
Unlike the case of energy, the law is affected by rotation and becomes (see Appendix A)
\begin{equation}
 \varepsilon_K = 2 \left\langle \delta \left( \uu \times \bOmega_T \right)  \cdot \delta \bomega \right\rangle   
+ 4 \left\langle (\uu \times \bomega) \cdot \bOmega \right\rangle   \, . \label{khrot}
\end{equation}
An interesting question is about the relative importance of the two terms appearing in this law. Direct numerical simulations can 
be used to check if one term is dominant and if this conclusion depends on the amplitude of the Rossby number \cite{mininni10a}.
In the limit of fast rotation (small Rossby number) we fall in the regime of weak wave turbulence for which a perturbation theory 
can be realized \cite{galtier03a,galtier2014}. It would be interesting to see, with the help of Eq. (\ref{khrot}), how the regimes of strong 
and weak wave turbulence, as well as the transition between them, are characterized when kinetic helicity is injected into the system.

\section{Incompressible MHD turbulence}

MHD turbulence is pervasive in astrophysics where ranges of scales available are huge \cite{biskamp2003}. In the case of the solar 
wind, {\it in situ} measurements are accessible and power law spectral distributions are commonly reported for the velocity and 
magnetic field fluctuations \cite{Galtier2006l,kiyani2015}. Like for neutral fluids, rigorous results are not only needed to better understand 
the physics of MHD turbulence \cite{Verma2004} 
but also to answer specific questions like the place and the amount of local heating that occurs in the solar wind where data clearly 
shows that the cooling of the wind with the heliospheric distance is slower than an adiabatic cooling \cite{Marsch1982}. 
Turbulence is believed to be a fundamental ingredient in this process because it furnishes naturally a source of local heating 
through the direct cascade and eventually the heating at small-scales. Exact relations can help in this problem because one is 
able to measure the field fluctuations with a spacecraft and then obtain the rate of energy dissipation (\ie the heating) \cite{luca}. 
However, space plasmas are not isotropic \cite{Matthaeus90,Stawarz09,Osman11} 
and therefore it is not easy to use differential relations like Eq. (\ref{vec1}) which necessitates multi-point measurements. 
The derivation of an alternative formulation of exact relations free of the divergence operator is therefore very welcome. 

Following the same methodology as above, we shall derive an alternative formulation for the exact relations in standard and Hall 
MHD turbulence. Hall MHD is a useful simple fluid model when one wants to investigate plasma scales smaller than the ion inertial 
length $d_i$ at which the ions and electrons become dynamically decoupled \cite{Stawarz2015}. For example, in the solar wind 
$d_i \sim 100$\,km. In this section we first derive the exact law corresponding to the total (kinetic + magnetic) energy conservation in 
Hall MHD, and then consider the second inviscid and ideal invariant of standard MHD, \ie the cross-helicity\footnote{For new exact relations 
corresponding to the magnetic and generalized helicity conservation in Hall MHD see \cite{Banerjee16}.} (unlike energy, cross-helicity 
is no longer conserved in Hall MHD).

\subsection{Conservation of total energy} \label{energymhd}
For convenience, the magnetic field will be normalized to a velocity. Then, the incompressible Hall MHD equations read 
\begin{align}
\p_t{\uu} &= - \nabla P_T + \uu \times \bomega + \jj \times \bb + \nu \nabla^2 \uu + \fz\, , \label{first1}\\
\p_t{\bb} &= \nabla \times \left[ \left(\uu - d_i \jj \right) \times \bb \right] + \eta \nabla^2 \bb + {\bf f_b} \, , \label{first2}\\
\nabla \cdot \uu &= 0 \, , \\
\nabla \cdot \bb &= 0 \, ,
\end{align}
where $\jj = \nabla \times \bb$ denotes the current density, $P_T=P+ u^2/2$ the total pressure, $\eta$ the magnetic diffusivity, 
$\fz$ and ${\bf f_b}$ are stationary forcing terms. In order to obtain an exact relation for the total energy conservation, we construct 
(similarly as the hydrodynamic case) the two-point energy correlators as 
\begin{equation}
R_{E} = R'_{E} = \left\langle \frac{\uu \cdot \uu' + \bb \cdot \bb' }{2} \right\rangle \, . 
\label{corrmhd}
\end{equation}
We will implicitly assume that $\uu$ and $\bb$ are fluctuating fields (\ie $\langle \uu \rangle = 0$ and $\langle \bb \rangle = 0$). 
Then, evidently we calculate the evolution of the correlators
\begin{align}
\p_t  &\left( R_{E} + R'_{E} \right) = \left\langle \uu' \cdot \p_t \uu + \uu \cdot \p_t \uu' + \bb' \cdot \p_t \bb + \bb \cdot \p_t \bb'  \right\rangle \nonumber \\
&= \left\langle \uu' \cdot \left[ - \nabla P_T + \uu \times \bomega  + \jj \times \bb \right] + 
\uu \cdot \left[ - \nabla' P'_T + \uu' \times \bomega' + \jj' \times \bb' \right] \right\rangle \nonumber \\
&+ \left\langle \bb' \cdot \left[  \nabla \times \left( \left( \uu - d_i \jj \right) \times \bb \right) \right] + \bb \cdot \left[  \nabla' \times \left( \left( \uu' - d_i \jj' \right) \times \bb' \right)\right] \right\rangle + {\cal D}_m + {\cal F}_m \nonumber \\
&= - \left\langle \delta \left[ \uu \times \bomega + \jj \times \bb \right] \cdot \delta \uu \right\rangle 
+ 2 \left\langle (\jj \times \bb) \cdot \uu \right\rangle \nonumber \\
&- \left\langle \delta ((\uu -d_i \jj) \times \bb) \cdot \delta \jj \right\rangle + 2 \left\langle ((\uu - d_i \jj) \times \bb) \cdot \jj \right\rangle 
+ {\cal D}_m + {\cal F}_m \nonumber \\
&=  - \langle \delta \left[ \uu \times \bomega + \jj \times \bb \right] \cdot \delta \uu \rangle 
- \langle \delta \left[\left(\uu - d_i \jj \right) \times \bb \right] \cdot \delta \jj \rangle + {\cal D}_m + {\cal F}_m \, ,
\end{align}
with the dissipation ${\cal D}_m = \nu \left\langle \uu' \cdot \nabla^2 \uu + \uu \cdot \nabla'^2 \uu' \right\rangle + 
\eta \left\langle \bb' \cdot \nabla^2 \bb + \bb \cdot \nabla'^2 \bb' \right\rangle$ and the forcing 
${\cal F}_m = \left\langle \uu' \cdot \fz + \uu \cdot \fz' + \bb' \cdot {\bf f_b} + \bb \cdot {\bf f'_b} \right\rangle$. 
For a statistical stationary state, and in the limit of a large-scale forcing, small viscosity and small magnetic diffusivity, 
we finally obtain the following exact relation for Hall MHD turbulence valid in the inertial range 
\begin{equation}
 2 \varepsilon_T = \langle \delta \left[ \uu \times \bomega + \jj \times \bb \right] \cdot \delta \uu \rangle + \langle \delta \left[\left(\uu - d_i \jj \right) \times \bb \right] \cdot \delta \jj \rangle \, , \label{hallfinal}
\end{equation}
where $\varepsilon_T$ is the mean rate of total energy dissipation (per unit of mass). Note that this relation is analogous to the one 
derived in \cite{Galtier08}, but here the final form is expressible purely in terms of two-point fluctuations. 
For length scales larger than the ion inertial length (it is equivalent to take $d_i \to 0$), we find the standard MHD regime and the 
corresponding exact relation is simply given by 
\begin{equation}
2 \varepsilon_T =  \langle \delta (\uu \times \bomega) \cdot \delta \uu \rangle + \langle \delta(\jj \times \bb) \cdot \delta \uu \rangle + \langle \delta(\uu \times \bb) \cdot \delta \jj \rangle
\end{equation}
which is nothing but an alternative form of \cite{Politanob98}. In the limit of very small-scales ($\ell \ll d_i$), we recover practically 
the inertialess electron MHD limit for which Eq. (\ref{hallfinal}) gets simplified to
\begin{equation}
 2 \varepsilon_T = - d_i \langle \delta \left( \jj \times \bb \right) \cdot \delta \jj \rangle \, . \label{elecmhd} 
\end{equation}
The law (\ref{elecmhd}) is interesting for solar wind turbulence because it can be verified accurately using single spacecraft data. Although it
includes the current density $\jj$, which is defined as the curl of the magnetic field and therefore involves spatial gradient (and hence the 
multi-spacecraft measure), we can in fact calculate $\jj$ using its alternative definition as 
\be
\jj = n e (\uu_i - \uu_e) \, , 
\ee
where n is the ion or electron number density, e is the protonic charge and $\uu_{i, e}$ denote respectively the ion and electron fluid velocities. All these
quantities for the solar wind will be soon available from the current Magnetospheric MultiScale (MMS/NASA) mission. 
\ADD{In fact, this space mission will produce very high resolution (burst modes) plasma data ($\sim 150 ms$ for the ions and $\sim 30 ms$ for the electrons) 
for only short periods of time of only $10$ to $20$ minutes. Only few moderate samples of $5$--$8 \, 10^3$ points will be available to evaluate the proton 
and electron velocities, the plasma density and finally, by deduction, the current density $\jj$. 
The question of statistical convergence is therefore crucial in this case to 
get a precise estimate of the transfer rate $\varepsilon_T$. Our new formulation in terms of second-order moment -- instead of third-order moment -- provides 
therefore an interesting alternative to measure accurately $\varepsilon_T$}, 
which would give a reliable measure of the turbulent heating rate of the solar wind thereby addressing one of the fundamental questions on space plasmas.

\paragraph{Effect of a background magnetic field} 
It is worthwhile to study the effect of a uniform background magnetic field $\bb_0$ on the energy exact relation (\ref{hallfinal}) as very often space 
plasmas are embedded in such a directive magnetic field.
Physically it is also relevant as the total energy is an inviscid invariant of Hall MHD even under the presence of $\bb_0$.
In this situation, the governing MHD equations become
\begin{align}
\p_t{\uu} &= - \nabla P_T + \uu \times \bomega + \jj \times (\bb + \bb_0) + \nu \nabla^2 \uu + \fz \, , \\
\p_t{\bb} &= \nabla \times \left[ \left(\uu - d_i \jj \right) \times (\bb+\bb_0) \right] + \eta \nabla^2 \bb + {\bf f_b} \, . 
\end{align}
The two-point energy correlators are still given by relation (\ref{corrmhd}) (with the assumption $\langle \uu \rangle = 0$ and $\langle \bb \rangle = 0$). 
We obtain
\begin{align}
\p_t  &\left( R_{E} + R'_{E} \right) = \left\langle \uu' \cdot \p_t \uu + \uu \cdot \p_t \uu' + \bb' \cdot \p_t \bb + \bb \cdot \p_t \bb'  \right\rangle \nonumber \\
&= \left\langle \uu' \cdot \left[ \jj \times \bb_0 \right] + \uu \cdot \left[ \jj' \times \bb_0 \right] \right\rangle \nonumber \\
&+ \left\langle \bb' \cdot \left[  \nabla \times \left( \left( \uu - d_i \jj \right) \times \bb_0 \right) \right] 
+ \bb \cdot \left[ \nabla' \times \left( \left( \uu' - d_i \jj' \right) \times \bb_0 \right)\right] \right\rangle + OT \nonumber \\
&= \left\langle \uu' \cdot \left[ \jj \times \bb_0 \right] \right\rangle + \left\langle \uu \cdot \left[ \jj' \times \bb_0 \right] \right\rangle \nonumber \\
&+\left\langle \left[(\uu - d_i \jj) \times \bb_0 \right] \cdot \jj' \right\rangle 
+ \left\langle \left[(\uu' - d_i \jj') \times \bb_0 \right] \cdot \jj \right\rangle \nonumber + OT \\
&= OT  \, ,
\end{align}
where $OT$ corresponds to the other terms derived previously. 
Here we see that the presence of a background magnetic field will not modify the analytical form of the law (\ref{hallfinal}). 
However, it is well known that it has a strong impact on the nonlinear dynamics with a reduction of the cascade along $\bb_0$ 
\ADD{(which can be understood by invoking simply the non-invariance of the MHD equations by a Galilean transformation with regards to $\bb_0$). 
This apparent inconsistency is solved if we push the analysis to the next order: indeed, as discussed in \cite{Oughton13} the equation for the time 
evolution of the third-order correlation depends explicitly on $\bb_0$, and hence so does the time-scale associated. Then, the second-order 
correlation depends implicitly on $\bb_0$, which may have therefore an impact on the nonlinear dynamics.}

\subsection{Cross-helicity conservation}
The second invariant of standard MHD is the cross-helicity whose density is given by $H_C = \uu \cdot \bb$. 
(Note that it is not an invariant of Hall MHD.)
The derivation is similar to that of the preceding subsection and thus we just give the result of our calculation, 
\begin{equation}
2 \varepsilon_C = \langle \delta (\uu \times \bomega) \cdot \delta \bb \rangle + \langle \delta(\jj \times \bb) \cdot \delta \bb \rangle 
+ \langle \delta (\uu \times \bb) \cdot \delta \bomega \rangle \, , \label{cross}
\end{equation}
where $\varepsilon_C$ is the mean rate of cross-helicity dissipation (per unit of mass). 
The conditions of application of this alternative law are the same as above. 
However, unlike Eq. (\ref{hallfinal}), the presence of a uniform background magnetic field $\bb_0$ alters the exact relation (\ref{cross}) (see Appendix B).

Interestingly, in contrast to the pure hydrodynamic case where the turbulent cascade vanishes as a result of the single alignment of $\uu$ and $\bomega$, 
MHD turbulence needs three alignments $(\uu, \bomega), (\bb, \jj) \text{ and } (\uu, \bb)$ for the turbulent cascade to vanish identically\footnote{Of course 
the net flux may vanish non trivially in many ways.}.
Finally, note that the introduction of the traditional Elsasser fields, $\uu \pm \bb$, does not lead to a simple expression (this is also 
true for the total energy) and therefore we do not use it.

\section{Conclusion}

In this paper we have derived an alternative form for some exact relations in the framework of homogenous (not necessarily isotropic) hydrodynamic 
and MHD turbulence. The new forms are free of the divergence operator and are therefore well adapted to anisotropic turbulence that is found 
naturally in a fluid under rotation or in space plasmas where a background magnetic field is present. 
\ADD{Unlike the classical exact relations, our alternative formulations involve only second-order correlations which should lead to a faster statistical 
convergence. This property may become crucial in some cases like for the measure of the heating rate in the solar wind.}
The current formulation presents the turbulent cascade as a deviation of Beltrami-type alignments. 
We also see that the Lamb vector ($\bomega \times \uu$) plays a key role in the energy and kinetic helicity cascade. Likewise, for MHD turbulence the 
Lamb vector and two additional vectors $(\uu \times \bb)$ and $(\jj \times \bb)$ play the central role in the cascading process for both energy and cross-helicity. 
In addition, by virtue of Eq. (\ref{elecmhd}), one can directly measure the energy dissipation rate in space plasmas at sub-ion scales only by 
using single spacecraft data as here we only need $\bb$ and $\jj$ for evaluating $\varepsilon_T$. 
A precise measurement of this quantity is important in collisionless plasmas where the fate of the energy is not well known since the dissipation
is likely to occur in a range of scales \cite{Told2015}. Therefore, the evaluation of $\varepsilon_T$ at different scales may give a strong constraint to theoretical models.

\section*{Acknowledgments}
Financial supports from PNST (Programme National Soleil--Terre) INSU--CNRS are acknowledged. 

\section*{Appendix A. Effects of a background rotation in hydrodynamics} 

In this appendix we derive the exact law for the kinetic helicity conservation (hydrodynamic turbulence) when a uniform background rotation is considered. 
The governing inviscid equations are then
\begin{align}
\p_t \uu &= - \nabla P_T  + \uu \times {\boldsymbol \bOmega_T} + \nu \nabla^2 \uu  + \fz \, , \\
\p_t \bomega &= \nabla \times \left( \uu \times \bOmega_T  \right) + \nu \nabla^2 \bomega + {\bf f}_{\bomega}  \, , 
\end{align}
where $P_T$ is the total pressure including the centrifugal term, $\bOmega_T = 2 \bOmega + \bomega$ where {$\bOmega$ is the constant rotating rate. 
As before, we will implicitly assume that there is no background velocity field. We obtain
\begin{align}
\p_t   &\left( R_{K} + R'_{K} \right) = \left\langle \uu' \cdot \p_t \bomega + \bomega \cdot \p_t \uu'  +  \uu \cdot \p_t \bomega' + \bomega' \cdot \p_t \uu  \right\rangle \nonumber \\
&= \left\langle \left[ - \nabla P_T + \left( \uu \times \bOmega_T \right) \right] \cdot \bomega' + \left[ \nabla' \times \left( \uu' \times \bOmega_T' \right) \right] \cdot \uu \right\rangle  \nonumber \\
&+ \left\langle  \left[ - \nabla' P_T' + \left( \uu' \times \bOmega_T' \right) \right] \cdot \bomega + \left[ \nabla \times \left( \uu \times \bOmega_T \right) \right] \cdot \uu'  \right\rangle + {\cal D}_K + {\cal F}_K \nonumber \\
&= 2 \left\langle \left( \uu \times \bOmega_T \right) \cdot \bomega' + 
\left( \uu' \times \bOmega_T' \right) \cdot \bomega \right\rangle + {\cal D}_K + {\cal F}_K \nonumber \\
&= - 2 \left\langle \delta \left( \uu \times \bOmega_T \right)  \cdot \delta \bomega \right\rangle   
- 4 \left\langle (\uu \times \bomega) \cdot \bOmega \right\rangle
+ {\cal D}_K + {\cal F}_K
\, . 
\end{align}

\section*{Appendix B. Effect of background magnetic field on the cross-helicity} 
Unlike the case of total energy, the presence of a uniform background magnetic field $\bb_0$ alters the exact relation for the cross-helicity conservation. 
Following the same methodology as above, we can show that 
\begin{equation}
2 \varepsilon_C = OT_C + \left\langle \left[ \left(\bomega \times {\bf u}' \right) + \left(\bomega' \times {\bf u} \right) + \left(\bb \times \jj' \right) + \left(\bb' \times \jj \right) \right] \cdot \bb_0 \right\rangle \, , \label{finalcross}
\end{equation}
where $OT_C$ represents the right hand side term in Eq. (\ref{cross}).

\section*{References}
\bibliography{banerjee_physicaD_7}
\end{document}